# Characterization of the strain rate sensitivity of basal, prismatic and pyramidal slip in Zircaloy-4 using micropillar compression

Ning Fang[1], Yang Liu[1], Finn Giuliani[1] and Ben Britton[1,2]*

1. Department of Materials, Imperial College London, Prince Consort Road, London, UK, SW7 2AZ

2. Department of Materials Engineering, University of British Columbia, 309-6350 Stores Road, Vancouver, BC Canada V6T 1Z4

*corresponding author: ben.britton@ubc.ca

## Graphical Abstract

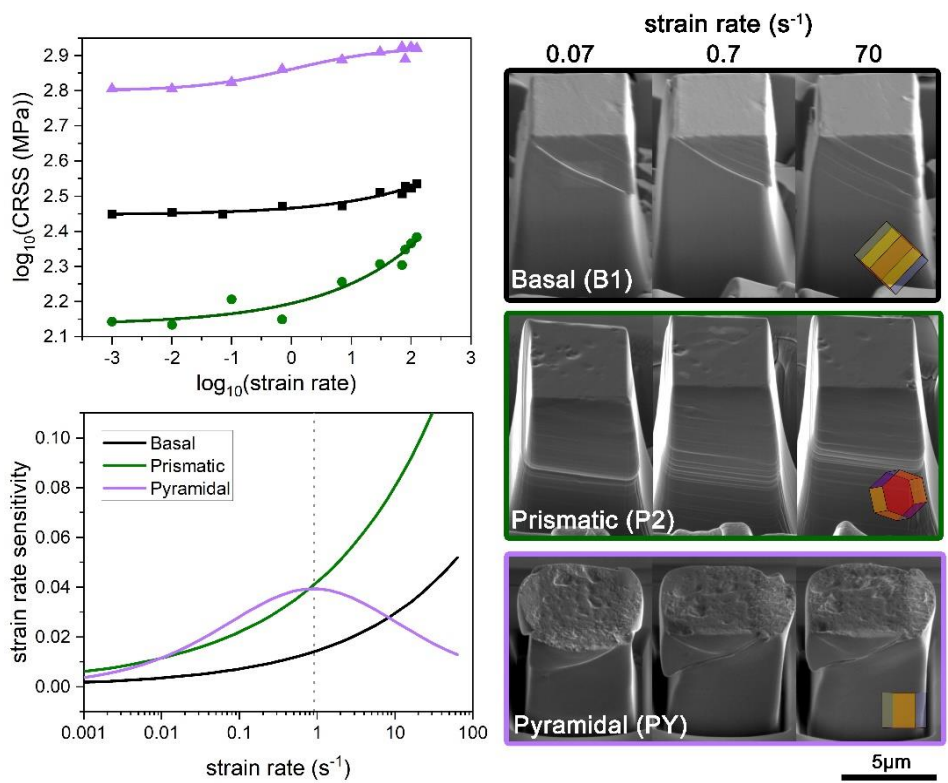

**Abstract**

The slip strength of individual slip systems at different strain rates will control the mechanical response and strongly influence the anisotropy of plastic deformation. In this work, the slip activity and strain rate sensitivity of the <a> basal, <a> prismatic, and <c+a> pyramidal slip systems are explored by testing at variable strain rates (from $10^{-4}$ $s^{-1}$ to 125 $s^{-1}$) using single crystal micropillar compression tests. These systematic experiments enable the direct fitting of the strain rate sensitivities of the different slips using a simple analytical model and this model reveals that deformation in polycrystals will be accommodated using different slip systems depending on the strain rate of deformation in addition to the stress state (i.e. Schmid's law). It was found that the engineering yield stress increases with strain rate, and this varied by slip systems. Activation of the prismatic slip system results in a high density of parallel, clearly discrete slip planes, while the activation of the <c+a> pyramidal slip leads to the plastic collapse of the pillar, leading to a 'mushroom' morphology of the deformed pillar. This characterization and model provide insight that helps inform metal forming and understanding of the mechanical performance of these engineering alloys in the extremes of service conditions.

1. Introduction

Zircaloy-4 (Zr4, Zr-1.5Sn-0.2Fe-0.1Cr, wt%) is typically used in the nuclear industry because of its low thermal neutron absorption cross-section. It is commonly used as fuel cladding and structural components in water-cooled nuclear reactors as it has excellent radiation stability and resistance to galvanic corrosion [1–5]. However, at room temperature, it is a transition metal with a hexagonal close-packed (hcp) crystal structure, and this leads to significant anisotropy in mechanical performance. Understanding this anisotropic mechanical performance is important when these alloys are used in demanding applications [6]. Specifically, for Zircaloy-4 the processing of cladding tubes is typically via cold pilgering which can result in a variation of strain rates that extend up to an order of $10^2$ s$^{-1}$ during forming of the thin tubes [7] and higher strain rates can be present in the extreme service conditions.

One way of considering deformation in materials is to focus on the idea that each small volume inside the material will change shape to reduce the energy of the system, via the easiest route to activate a deformation pathway. The ability for these pathways to activate can be controlled by the availability of existing defects to move (e.g., vacancies, dislocations, and interfaces such as twinning and grain-boundaries or cracks) and new defects to be nucleated and then move/propagate. Many of these defect nucleation or propagation processes are typically thermally activated (e.g. the movement of dislocations via jogs, kinks and point defects) and therefore there is an inherent strain rate sensitivity for each of these mechanisms. In practice, in some material systems the strain rate sensitivity can be quite low and is often ignored and yet in others, such as zirconium-based alloys, the strain rate sensitivity can be quite large for both dislocation slip and the competition between twinning and slip

Materials with anisotropic slip systems and potential for deformation twinning (such as the hcp metals), can be affected by the relative strain rate sensitivity of these different deformation modes. Furthermore, as crystallographic reorientation is often controlled or mediated by deformation (twinning and lattice rotation due to constrained slip) and associated structures (e.g. via recovery and recrystallization), the relative strain rate sensitivity will also control or be controlled by the crystallographic texture of the polycrystalline aggregate.

Although twinning plays a more important role at high strain and temperature in many metals, in zirconium-based alloys the volume percentage of twinning is rarely significant for strains less than 0.1 and at room temperature, according to both experimental and simulation results [8,9]. This motivates our focus in the present study on the strain rate sensitivity of the slip systems only.

Understanding the relative strain rate sensitivity for individual mechanisms, especially at strain rates > $10^2$ $s^{-1}$ is extremely challenging experimentally, as conventional experiments require specimens typically to be large (>1 $mm^3$) and this results in an indirect measurement of stress-strain response of polycrystalline samples (e.g. via the split Hopkinson pressure bar, cam plastometer and drop test, and Taylor impact test). As an indicator of the challenges, to perform a high-quality high rate test, the analysis also needs to take into account a variety of dynamic factors including heat effects, wave propagation effects, shock wave effects, and inertia effects [10–12].

For materials like Zircaloy-4, where single crystal growth is challenging, extraction of single crystal specimens for variable strain rate testing is prohibitively expensive and so indirect extraction of properties from (textured) polycrystalline aggregates is required.

To address the issues associated with the indirect extraction of single slip system properties from polycrystals, micropillar compression can be used. Micropillars are cut from polycrystalline materials using focused ion beam (FIB) machining to extract single crystal specimens which are subsequently compressed using a small-scale mechanical tester (e.g. a nanoindentation system equipped with a flat punch), and the external geometry and size of the micropillars are significant for the characterization of the material property [13].

For different test parameters, micropillar compression tests have been applied in many areas with the development in nanoindentation instrumentation [14], such as tests under high cycle fatigue [15] and various temperatures [1,16–19]. For cases where these tests reveal the activity of more than one slip system, Li *et al.* [20] explained the prismatic-to-basal plastic slip transition using the theory of mobility laws and prismatic-to-basal cross-slip energy barrier.

For high strain rate tests, apart from the metals mentioned above, there have been studies to extend high rate micropillar compression as a tool to understand deformation in nanocrystalline glasses [21], ceramics [16] and polymers [22].

For materials with hcp structure, a lot of work with micropillar compression has already been done, e.g. for Mg [16,23–27], Ti [28–32] and Zr [1,18,33,34] but these studies have been largely quasistatic or with slower strain rates (below $10^{-2}$ s$^{-1}$). However, even if metals are in the same group, the mechanism, especially the strength, hardness and strain rate sensitivity of their alloys differ considerably due to the manufacturing process and the properties of the metals themselves. For example, Ventura *et al.* [35] highlight the importance of twinning mechanisms during the compression of magnesium, which is caused by the significant slip anisotropy and bonding type. Titanium alloy also shows a different relationship among different slip systems and strain rate sensitivity [34,39], compared with zirconium alloy.

In the previous work, Gong *et al.* [36] measured that the CRSS ratio for their samples is approximate <a> prismatic : <a> basal : <c+a> 1$^{st}$ pyramidal = 1:1.3:3.5 at quasi-static strain rates, as reported for commercially pure zirconium. In most experiments, <c+a> pyramidal slip is extremely hard to isolate and characterize, as it has a very high critical resolved shear stress and therefore small misalignment of the mechanical test or realignment during testing, will result in deformation largely being accommodated by other mechanisms.

In this paper, micropillar compression deformation has been performed in five selected grains at different strain rates to explore the variation in slip strength and the difference in the slip traces on the sides of each micropillar. An analysis of the strain rate sensitivity based on the individual slip system and multiple slips was carried out and combined with an analytical model. The objective of this work is to explore the relationship between different slip systems and predict slip strength and slip activity for varying strain rates.

## 2. Experimental procedure/ Methods

### 2.1. Grain characterization and selection

Zircaloy-4 (Zr4, Zr-1.5Sn-0.2Fe-0.1Cr,wt%) samples with 'blocky alpha' were heat-treated for 336 hours (14 days) at 800 °C in an argon atmosphere (following the recipe developed by Tong & Britton [37]), and this results in a sample with large grains (typically >500 µm). The large grain samples were metallographically ground up to a 2400 grit SiC finish, followed by broad ion beam polishing using a PECS II (Gatan, Inc. Pleasanton, USA) for 15 minutes at room temperature (following the optimized recipe of Fang *et al.*[38]). The settings of the broad ion beam are 8 keV, 8° beam angle, no modulation and 1 rpm.

To select grains that have a high Schmid factor for the operation of one system in preference to any other, the orientations of the grains in the sample were obtained using electron backscatter diffraction (EBSD) analysis in a Quanta 650 FEG SEM with a Bruker eFlash HR camera. EBSD mapping was carried out with an ~10 nA focused electron beam operating at 20 kV and with a working distance of 16.8 mm. The EBSD detector was inserted into the chamber at a detector distance of 17 mm and 10.62° tilt. The EBSD maps were collected with a step size of 11.6 µm with 400 x 266 points per map and patterns were collected with 200 x 200 pixel resolution EBSPs with an exposure time of 19.4 ms per pattern. The resultant bcf file was converted to .h5 files and processed with MTEX [39] in MATLAB (using the conventions as the direction of x axis pointing west, y axis pointing down and the direction of z axis pointing out of the plane).

Figure 1 shows an MTEX-processed EBSD mapping of heat-treated and well-prepared Zircaloy-4, showing the large 'blocky-α' grains and grain orientations concerning the sample surface (i.e., looking along the loading axis of cut micropillars along the Z-axis).

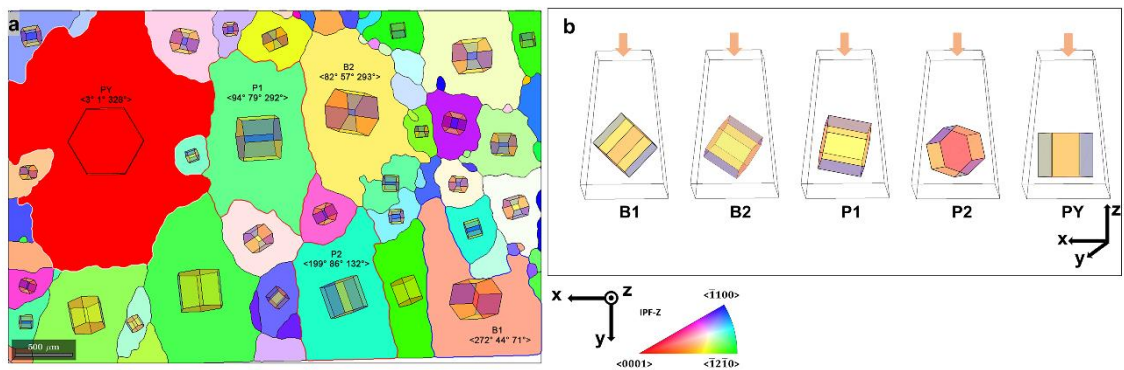

*Figure 1 (a) An EBSD mapping with IPF-Z colouring and crystal shapes of the region of interest in the 'blocky alpha' large grain Zircaloy-4. (b) A schematic of the micropillar with the hcp crystal shape representing its orientation inside.*

Among these grains, five grains (Figure 1(b)) were identified that enable mechanical testing where slip activity is expected predominantly on one slip system (as it has a high Schmid factor (SF) compared to the other slip systems, as shown in Table 1):

| | Largest Schmidt factor in each slip system | | | | |
|---|---|---|---|---|---|
| | B1 | B2 | P1 | P2 | PY |
| Orientation | <272° 44° 71°> | <82° 57° 293°> | <94° 79° 292°> | <199° 86° 132°> | <3° 1° 328°> |
| <a> basal | 0.4905 | 0.4534 | 0.1855 | 0.0681 | 0.0154 |
| <a> prism | 0.2389 | 0.3381 | 0.4675 | 0.4948 | 0.0001 |
| 1st <a> pyramidal | 0.3665 | 0.4283 | 0.4658 | 0.4570 | 0.0075 |
| 1st <c+a> pyramidal | 0.3933 | 0.2969 | 0.4806 | 0.3984 | 0.4134 |
| 2nd <c+a> pyramidal | 0.2821 | 0.3756 | 0.4897 | 0.4561 | 0.4567 |

*Table 1 List of the grain orientation and the largest Schmidt factor in each slip system for the five selected grains. [Detailed information on the Schmid factor calculation and values of each slip system can be found in the supplementary.]*

The grain B1 was identified for micropillar tests that have basal slip system with a large $SF_{Basal}$ close to 0.5, but the other two values of the three $SF_{Basal}$ differed significantly, and the larger one (0.3278) has the potential to be activated. This motivated the identification of B2, where the other two basal slip systems have a lower SF (0.1785 & 0.2749). However, for B2 the largest $SF_{prism}$ is 0.3381, and the largest $SF_{1st\ py<a>}$ is 0.4283. Based on the current knowledge of the critical resolved shear stress (CRSS) ratios of different slip systems, there is potential that prismatic may be activated for this orientation and this was noted when the post-test analysis was performed.

In the search for potential grains that could activate prismatic slip systems, P1 was found first, it has a large $SF_{prism}$ (0.4675), while values of $SF_{Basal}$ are all very small, but the values of $SF_{1st\ py<a>}$ and $SF_{prism}$ are close to each other. Therefore, another grain P2 was found, whose $SF_{prism}$ is 0.4948.

Finally, grain PY was identified which has the c-axis almost in parallel with the loading direction. This crystal orientation will not result in the activation of basal and prism. In its SF calculations, the values in all six directions of $SF_{2nd <c+a>py}$ are all close to 0.45, while $SF_{1st <c+a> pyramidal}$ are all close to 0.4.

## 2.2. Micropillar fabrication and compression

In each grain, the Ga-FIB based micropillar fabrication was performed using a Thermo Scientific™ Helios™ 5 CX DualBeam. An automated script in Thermo Scientific NanoBuilder was used for reproducible fabrication using a multi-step process: Milling was performed using four steps of 30 kV $Ga^+$ ion beam milling, with a reduction in each current for each cutting step: 21 nA was used to cut out a large area around the pillar; the pillar was gradually reduced using currents of 9 nA, 6.5 nA and finally to 0.79 nA. The target geometry of all the micropillars was at approximately 5 μm in width (middle), and 10 μm in height, and there was some variability due to the state of the $Ga^+$ ion beam (due to aperture wear etc between sessions) and the grain orientation. Therefore, the final dimensions of each fabricated pillar were measured using scanning electron microscopy (SEM) after each pillar was made in the same instrument.

Micropillar compression testing was carried out in the Quanta 650 FEG SEM using displacement control within two Alemnis nanoindentation systems. These indenters are displacement controlled, and two different frames are used to cover a wider range of strain rates, as shown in Figure 2. In this paper, the quasi-static module is referred to as the normal strain rate (NSR) testing and the high strain rate experiment is referred to HSR.

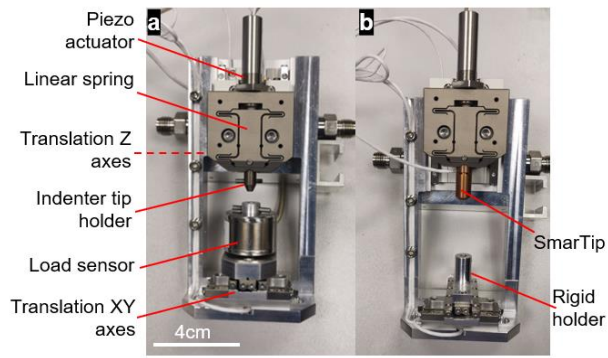

*Figure 2 The hardware components of the Alemnis nanoindenter stages for different strain rate testing (a) quasi-static, and (b) high strain rate.*

As the indenter tip displacement is driven by a piezoelectric crystal, the indenter tip can move extremely quickly in the NSR module, however the load cell cannot accurately capture the load response due to resonance. These so-called eigenfrequency resonances create substantial oscillations in the read out and affect the ability to perform tests above a high strain rate, which is higher than the load cell eigenfrequency which occurs at a displacement rate of 50 µm/sec.

To test at high strain rates, a second frame is used with a different mechanical load-train. In this frame, the load cell is replaced with a (near) rigid holder and a second capacitance-based sensor is mounted on the indenter tip, which is called a SmarTip (ST-025) and contains 1 axis of actuation + 3 axis sensors, using a piezoelectric crystal that measures accumulated displacements (which is calculated from the voltage) via electrodes and this is converted into stresses (similar to the internal operation of a load cell) using calibration and the supplied software 'HSR to XYZ'.

For the experiments in this work, the setting used to obtain reasonable displacement profiles which vary with time can be found in Table 2. More detailed settings can be found in Supplementary Information 1.

| Frame | Strain rate($s^{-1}$) | Load Speed (μm/s) | Load time(s) | Hold time (s) | Unload time (s) |
|---|---|---|---|---|---|
| NSR | 0.001 | 0.01 | 100 | 100 | 100 |
|  | 0.01 | 0.1 | 10 | 10 | 10 |
|  | 0.1 | 1 | 1 | 1 | 1 |
| Frame | Strain rate($s^{-1}$) | Load Speed (μm /s) | Output Frequency (Hz) | HSR Sampling Rate (Hz) | STD Sampling Rate (Hz) |
| HSR* | 0.07 | 1 | 1K | 1k | 1K |
|  | 0.7 | 10 | 10K | 10k | 10K |
|  | 7 | 100 | 100K | 100K | 50K |
|  | 30 | 500 | 1M | 200K | 50K |
|  | 70 | 1000 | 1M | 1M | 50K |
|  | 80 | 1300 | 1M | 1M | 50K |
|  | 100 | 1600 | 1M | 1M | 50K |
|  | 125 | 1900 | 1M | 1M | 50K |

*\* The shape of the voltage profile curve contains an exponential option; thus, the time-displacement curve is non-linear. This strain rate is calculated from the linear part during loading.*

*Table 2 Displacement profile of NSR testing and voltage profile input to the system and output sampling rate of high strain rate testing.*

The collection of HSR load-displacement data includes significant instrument noise that makes data analysis difficult. This motivates a systematic approach to reduce this noise in the data. The weighted regression method was used for the analysis presented within this paper as it was found to provide the most consistent and interpretable results, which relies on a smoothing function (Curve Fitting Toolbox in Matlab) using a weighted regression (via 'rloess'), where the filter assigns zero weight to data outside six mean absolute deviations. This method has less error and retains the original trend of the curve well.

Meanwhile, the difference in the hardware mainly leads to the difference in the stiffness system compressing the samples and this requires calibration (detailed calibration can be found in Supplementary Information 2) and careful data analysis to compare the two systems.

After capturing the surfaces of deformed micropillars, the sample surface is polished again to remove all the pillars to ensure the entire micropillars are located inside a grain, avoiding containing any grain boundaries.

## 3. Results

### 3.1. Uncertainty analysis

Due to the limitation of grain size and the requirement of certain intervals between micropillars, the number of micropillars that can be cut and tested simultaneously within the same grain is restricted.

In addition to repeating experiments, another two methods were employed to enhance the reliability of the experiments. Firstly, similar experimental results are compared, such as coupling the results of 0.07 $s^{-1}$ with 0.1 $s^{-1}$ in different micropillar compression regimes and using 70 $s^{-1}$ with 80 $s^{-1}$ to test similar experimental conditions in the same regime. Secondly, different grains (e.g. B1 and B2) were chosen to activate the same grain, although other slip systems were activated as shown in the results section.

Three sets of experiments were performed for each experimental condition. As shown in Figure 3, this is the case for strain rate 0.1 $s^{-1}$. It can be observed that there are fluctuations in the data points, and the range of fluctuations in test 1 and test 2 are relatively consistent. During the calibration process of different experimental parameters and stiffness, the error caused by data reading corresponding to 0.3% engineering strain was more evident than the error caused by fluctuations in the data fitting process.

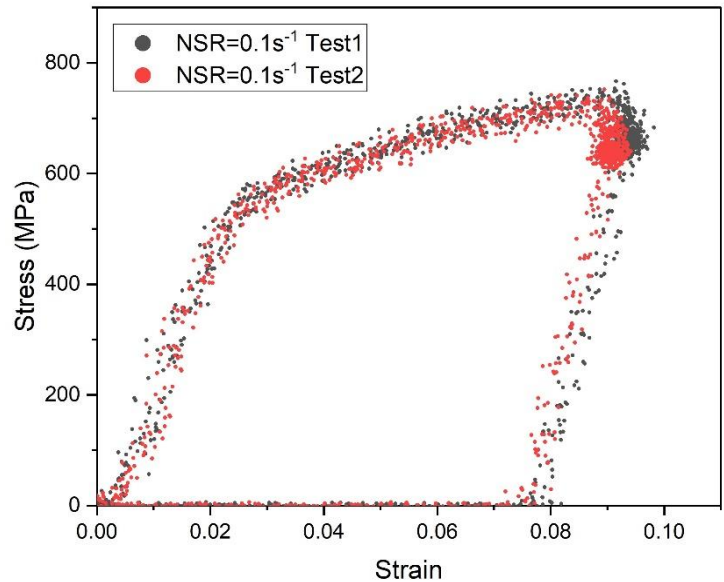

*Figure 3 Two micropillar compression tests on the pillars in B1 with quasi-static strain rates 0.1 $s^{-1}$.*

Therefore, to enable the analysis with more comprehensive error considered, especially at higher rates, a 0.3% offset yield stress was used for these calculations, and the errors are measured from the 0.2% offset yield stress and 0.4% offset yield stress.

### 3.2. B1 - <a> basal slip system

Figure 4(a) shows the stress-strain curves of micropillars in grain B1 among different strain rates. Data from the NSR and HSR testings are presented after smoothing, using the weighted 'rloess' curve fitting method, which can be found in Figure 4(b).

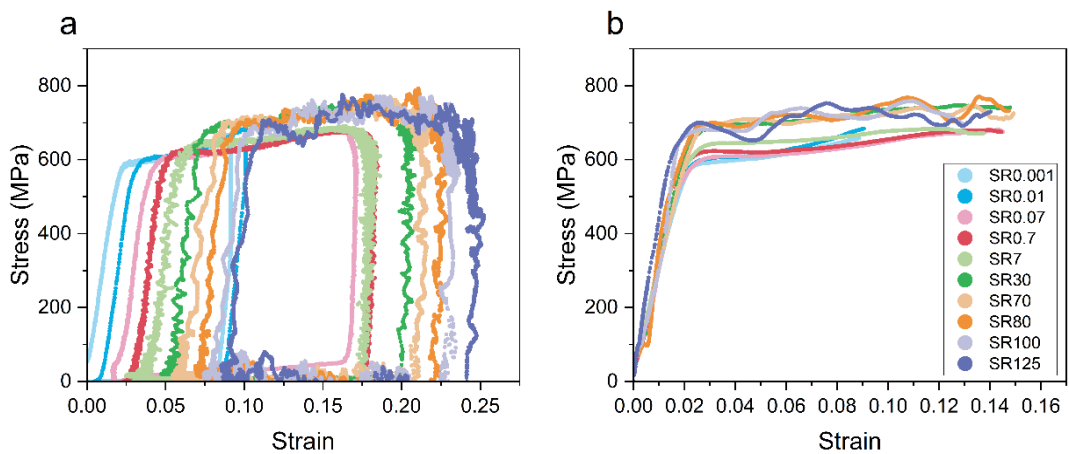

*Figure 4 (a) The coupled engineering stress-strain curves for micropillar compression tests in B1 at both NSR and HSR. [Each test has shifted slightly along the X axis to aid visualisation.] (b) Smoothed stress-strain curves of the coupled B1 results.*

When the micropillar shows yielding after elastic deformation, it reaches the proportional limit and starts to deviate from its linear slope. It can also be found that the hardening rate is larger at a low strain rate.

For the present work, a 0.3% offset yield strength is selected to characterize the yield of each slip system. The corresponding stress of ±0.1% strain (0.2% & 0.4% offset yield strength) is read to obtain the uncertainty error value.

After the compression, scanning electron (SE) imaging was performed on the four sides of the micropillars. Figure 5 shows an example from a micropillar cut in grain B1 and deformed with a strain rate of 0.7 s$^{-1}$.

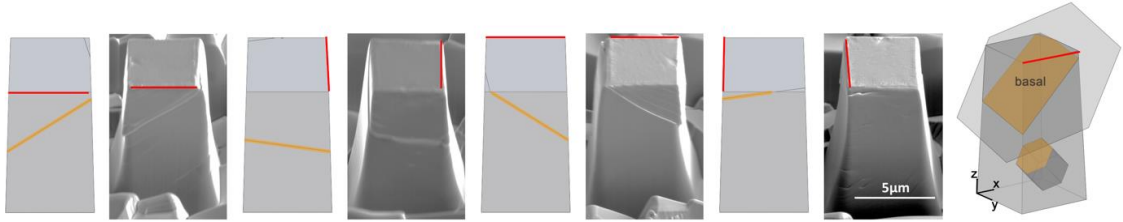

*Figure 5 SE micrographs and 3D models showing the relationship between pillars and slip planes for the four sides of the deformed pillar B1 at a strain rate of 0.7. [The red line annotation indicates the same pillar vertex for each micrograph.]*

The EBSD data and the (pre-deformation) SEM micrographs were used to generate a 3D drawing of the intersection of potential active slip planes and the pillar faces, which is shown in Figure 5. This assists in verifying observable slip traces.

A comparison of the 3D model and the post-deformation pillar faces in Figure 5 demonstrates the activation of the basal slip system, and the schematic and SEM images are well coupled. Small errors could be associated with the alignment of the identified traces (from the 3D model) and the experiments from these three sources.

- Uncertainty in the micropillar geometry and taper angle.

- During the deformation process, the pillar reduces in height and increases in width due to plastic deformation. Together with constraint due to the fiction of the pillar and the indenter tip, and that the pillar is attached to the based substrates, this can result in small rotations of the slip plane especially when the strain is large.

- Interaction between slip bands due to the activity of slip from more than one source and the activation of a second slip system (which is not apparent for this pillar).

Overall, slip trace analysis indicates that only B1 activates within this grain.

### 3.3. B2 – Combined <a> basal and <a> prismatic slip

Figure 6 shows the stress-strain curve of micropillars in grain B2 among different strain rates. Data from the NSR and HSR presented in smoothed data is shown in Figure 6(b). After the compression, SE imaging was performed on the four sides of the micropillars.

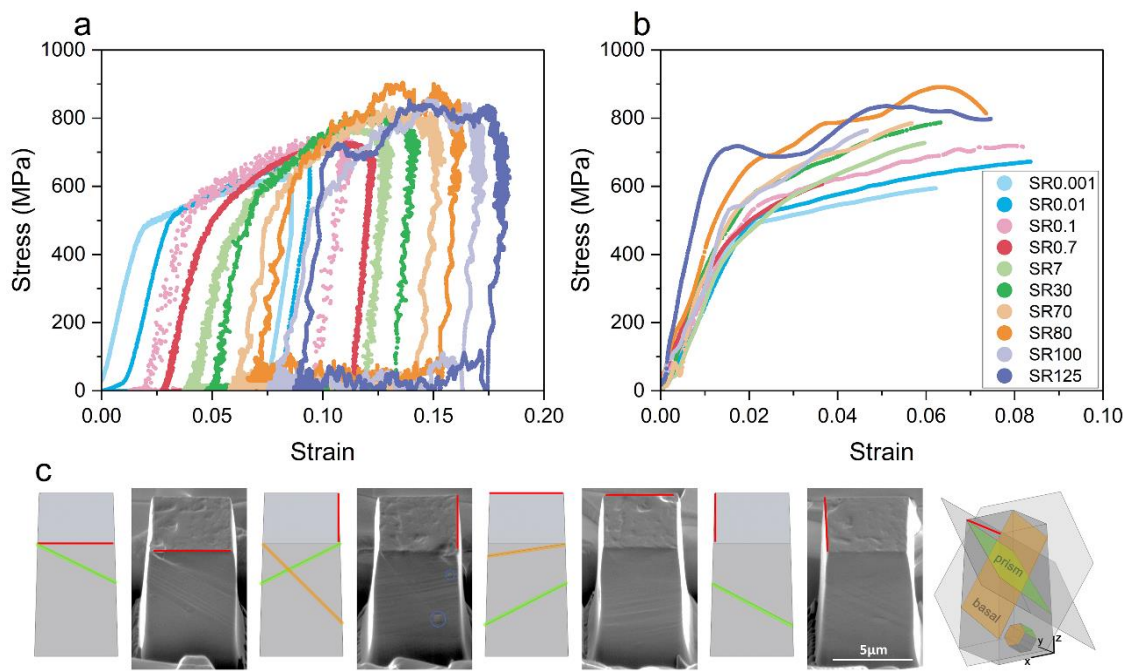

*Figure 6 (a) The coupled engineering stress-strain curves for micropillar compression tests in B2 at both NSR and HSR. [Each test has shifted slightly along the X axis to aid visualisation.] (b) Smoothed stress-strain curves of the coupled B2 results; (c) SE micrographs and 3D models showing the relationship between pillars and slip planes for the four sides of the deformed pillar B2 at a strain rate of 100. [The red line annotation indicates the same pillar vertex for each micrograph.]*

In contrast to B1, the 'nose' of the stress-strain curve in B2 is not that sharp, which means the yielding behavior includes significant microplasticity Furthermore, the elastic loading stiffness varies, even though the unloading stiffness has been calibrated. The hardening rate is increasing with a larger strain rate.

In the SEM imaging in Figure 6(c), it can be seen that the basal slip trace is shown as a single slip band, while the prismatic slip system appears as many slip traces in parallel.

The slip trace is evident on the side of pillars in the SE imaging, but not all the intersections between the micropillar and the slip plane can be found on each face. Only two basal slip traces can be found on the side of the pillar, which originates from a top corner, and another two invisible potential traces are not.

Some secondary phase particles (SPPs) can be seen on the side of the pillar (as shown in the blue circle in Figure 6(c)). In general, SPPs are commonly dispersed within Zircaloy-4 and are typically important for corrosion resistance but they could also impact slip. The parallel prismatic slip trace is not interrupted by these SPPs, and instead the slip trace shows that slip moves around the SPPs. This does not affect the angle of the projected slip trace on the side of the pillar, indicating that once the slip moved around the SPPs, deformation resumes a parallel slip plane from the same slip system.

### 3.4. P1 – the mixture of <a> prismatic and <a> pyramidal slip systems

Grain P1 was selected to activate <a> prismatic slip and to explore how <a> pyramidal slip might interact with this. The stress-strain curves of micropillars in grain P1 among different strain rates are shown in Figure 7(a). Curves from the NSR and HSR testing presented in smoothed data are shown in Figure 7(b). After the compression, SE imaging was performed on the four sides of the micropillars. Figure 7(c) shows an example from a micropillar cut in grain P1 and deformed with a strain rate of 125 s$^{-1}$.

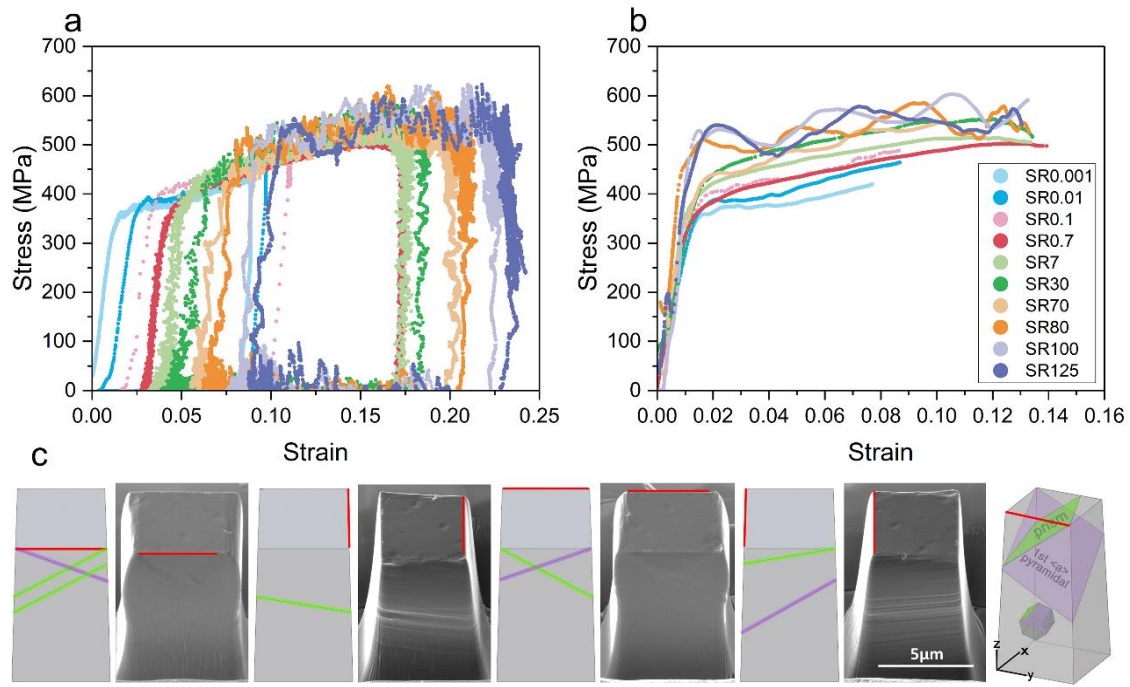

*Figure 7 (a) The coupled engineering stress-strain curves for micropillar compression tests in P1 at both NSR and HSR. [Each test has shifted slightly along the X axis to aid visualisation.] (b) Smoothed stress-strain curves of the coupled P1 results; (c) SE micrographs and 3D models showing the relationship between pillars and slip planes for the four sides of the deformed pillar P1 at strain rate 125. [The red line annotation indicates the same pillar vertex for each micrograph.]*

In the SE imaging in Figure 7(c), analysis of the slip traces indicates that there are many parallel prismatic slip bands. The schematic of the slip traces in the 3D model also includes the <a> pyramidal slip plane, as analysis of the slip near the top of the pillar revealed a slip trace that is consistent with this projected plane.

### 3.5. P2 – a mixture of two different prismatic slip systems

Another grain with $SF_{prism}$ = 0.4948 resulted in the activation of a prismatic slip system, and the experimental data from this grain is easier to interpret. Figure 8 shows the stress-strain curves of micropillars in grain P2 among different strain rates, together with the coupled and smoothed stress-strain curves, and the post-deformation SE imaging with a strain rate of 70 s$^{-1}$ in P2.

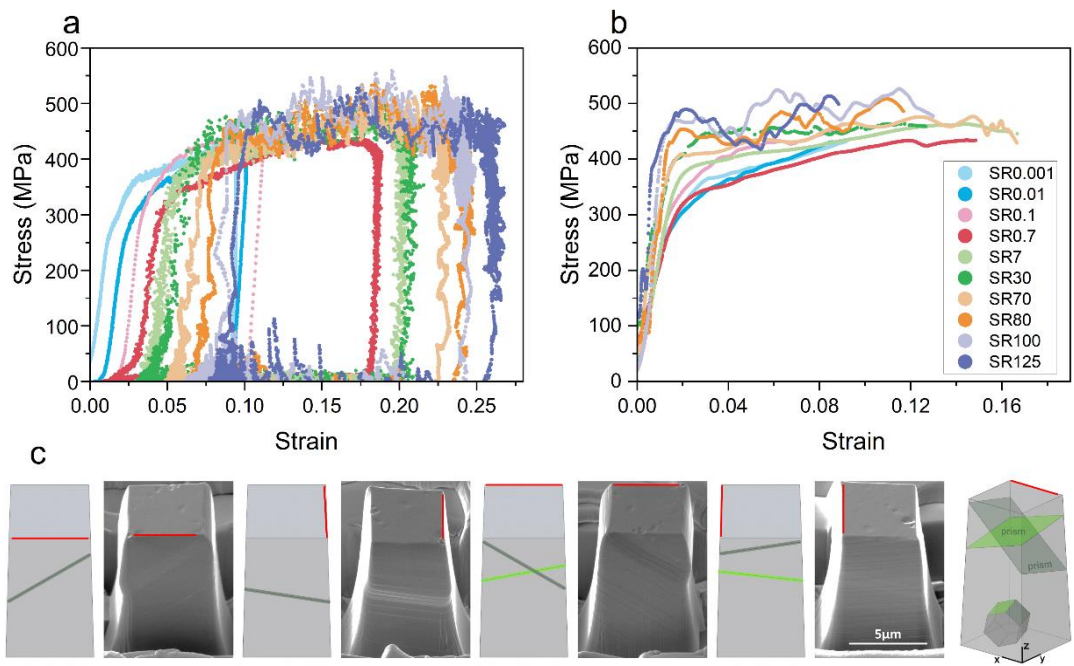

*Figure 8 (a) The coupled engineering stress-strain curves for micropillar compression tests in P2 at both NSR and HSR. [Each test has shifted slightly along the X axis to aid visualisation.] (b) Smoothed stress-strain curves of the coupled P2 results; (c) SE micrographs and 3D models showing the relationship between pillars and slip planes for the four sides of the deformed pillar P2 at strain rate 70. [The red line annotation indicates the same pillar vertex for each micrograph.]*

The 2$^{nd}$ slip system can be found in Figure 8(c), and it seems one group of parallel slip traces is 'beneath' the other group of parallel prismatic slip traces, and the 2nd slips can be found on three faces. From the reference of 3D micropillars, another prismatic (SF = 0.2925) slip system activates in preference to slip on an alternative and potential 1st <a> pyramidal (SF = 0.4570) slip system.

Analysis of the *in situ* SEM video was carried out from the NSR testing to further help understand slip activity during these tests. The surface of the pillars indicates that there is multiple <a> prismatic slip on parallel slip planes, and with increasing strain rates shown in Figure 9, the slip plane trace is found nearer to the top of the pillar, presumably related to the strain rate sensitivity of this deformation mode.

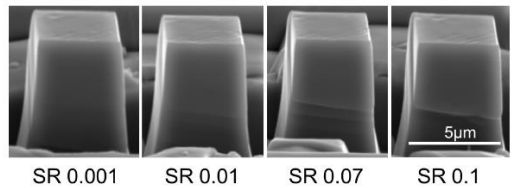

*Figure 9 The SE images of deformed micropillar in P2 with different quasi-static strain rates. (The in situ videos can be found in the supplementary figures.)*

### 3.6.   PY - 1st <c+a> pyramidal

A grain was identified where the loading axis is well aligned along the <c> direction, thus enabling analysis of <c+a> slip (as the <a> directions are poorly aligned).

Figure 10 shows the stress-strain curves of micropillars in grain PY for varying strains. Data from the NSR and HSR testing are shown in Figure 10(b). After the compression, SE imaging was performed on the four sides of each micropillar. Figure 10(c) shows an example from a micropillar cut in grain PY and deformed with a strain rate of 125 s$^{-1}$. For this crystal orientation, initial tests at quasi-static rates showed stress drops were observed at ~0.09 strain, and so later tests were conducted for strains of up to 0.2 to understand this phenomenon more.

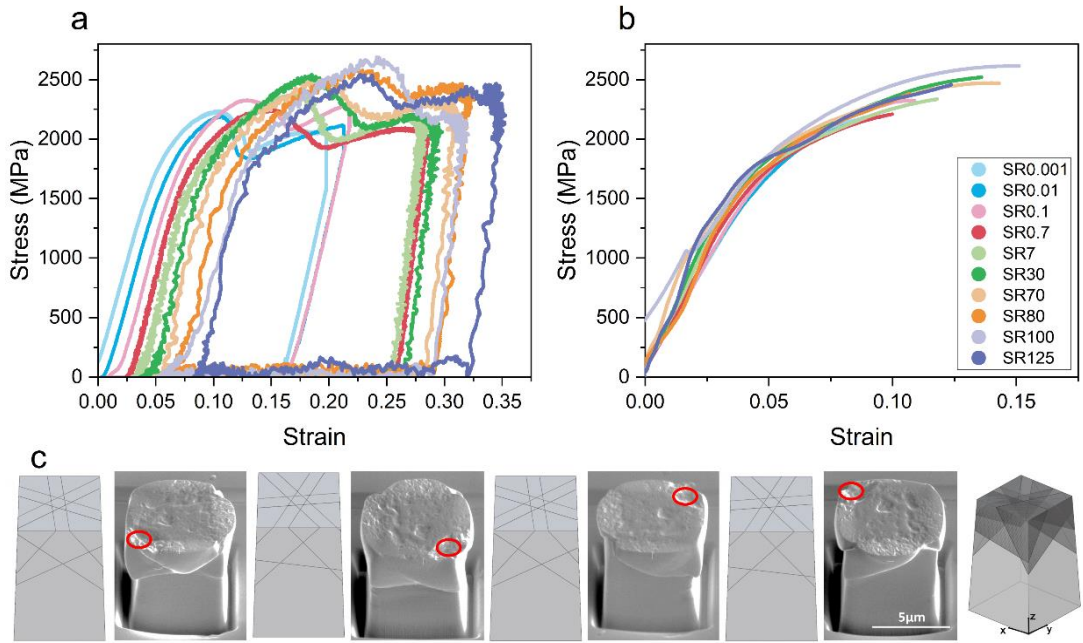

*Figure 10 (a) The coupled engineering stress-strain curves for micropillar compression tests in PY at both NSR and HSR. [Each test has shifted slightly along the X axis to aid visualisation.] (b) Smoothed stress-strain curves of the coupled PY results; (c) SE micrographs and 3D models showing the relationship between pillars and slip planes for the four sides of the deformed pillar PY at strain rate 125. [The red line annotation indicates the same pillar vertex for each micrograph.]*

This pillar orientation shows significantly different yielding and hardening behavior when compared to basal and prismatic slip systems (compare Figure 10, with Figure 4 and Figure 8). The transition from elastic to plastic deformation is at much higher stress (~1.5 GPa) as compared to the <a> slip systems, and this transition has a smooth shoulder. In the electron micrographs, these pillars show a significant collapse of the top of the pillar, and slip traces of these deformation structures are consistent with the predicted slip traces for <c+a> type slip. With the help of the 3D model, although the potential scenarios are more complex, it is possible to find out the 1st <c+a> pyramidal slip planes are mainly activated. For this set of slip systems, the Schmid factors are very close together for all these slip systems for this crystal orientation, and this means that multiple slip systems can operate during deformation to accommodate plastic strain.

## 4. Discussion

### 4.1. Slip strength vs. strain rate

Analysis of the critical resolved shear stress, evaluated using Schmid's law, has been performed and the results are given in Table 3 and plotted in Figure 11 (a-c). To enable a fair comparison, especially at higher rates, a 0.3% offset yield stress was used for these calculations.

| | Strain Rate | B1 | | B2 | | P1 | | P2 | | PY | |
|---|---|---|---|---|---|---|---|---|---|---|---|
| | | CRSS | error | CRSS | error | CRSS | error | CRSS | error | CRSS | error |
| NSR | 0.001 | 280.8 | -8.1/+3.7 | 213.1 | -11.3/+4.5 | 162.7 | -6.1/+4.2 | 139.0 | -3.0/+4.5 | 639.4 | -63.9/+36.5 |
| | 0.01 | 284.5 | -4.9/+3.4 | 195.0 | -40.8/+18.1 | 164.1 | -5.1/+8.9 | 136.1 | -6.9/+3.5 | 639.4 | -45.7/+37.4 |
| | 0.1 | NaN | NaN | 231.2 | -18.1/+14.1 | 179.5 | -6.5/+4.2 | 160.8 | -4.9/+4.9 | 666.8 | -45.7/+41.1 |
| HSR | 0.07 | 280.8 | -8.1/+4.7 | NaN | NaN | NaN | NaN | NaN | NaN | NaN | NaN |
| | 0.7 | 296.3 | -5.4/+2.9 | 217.6 | -10.0/+8.2 | 170.2 | -3.7/+2.8 | 141.0 | -11.9/+7.4 | 726.2 | -22.8/+25.1 |
| | 7 | 297.5 | -6.1/+5.2 | 195.0 | -18.1/+9.1 | 188.9 | -3.3/+2.8 | 180.1 | -5.9/+4.0 | 771.8 | -20.6/+13.7 |
| | 30 | 324.2 | -6.4/+4.4 | 266.6 | -5.9/+4.5 | 196.8 | -5.1/+4.2 | 202.9 | -1.5/+4.9 | 812.9 | -18.3/+13.7 |
| | 70 | 321.8 | -12.8/+10.1 | 266.6 | -5.9/+5.4 | 200.6 | -4.2/+1.4 | 201.4 | 0.0/+1.0 | 840.3 | -13.7/+9.1 |
| | 80 | 337.5 | -5.6/+2.9 | 313.8 | -0.9/+3.6 | 234.7 | -0.5/+2.8 | 222.7 | -0.5/+1.0 | 776.4 | -18.3/+11.4 |
| | 100 | 333.0 | -1.2/+1.0 | 263.0 | -9.1/+4.5 | 246.8 | -2.8/+0.9 | 231.6 | 0.0/0.0 | 840.3 | -18.3/+9.1 |
| | 125 | 342.9 | -1.2/+0.5 | 311.9 | -0.5/+0.9 | 252.5 | -0.9/0.0 | 241.5 | -2.5/+0.5 | 833.5 | -11.4/+6.9 |

*Table 3 CRSS value of different slip systems with varying strain rates from 0.001 to 125 s$^{-1}$.*

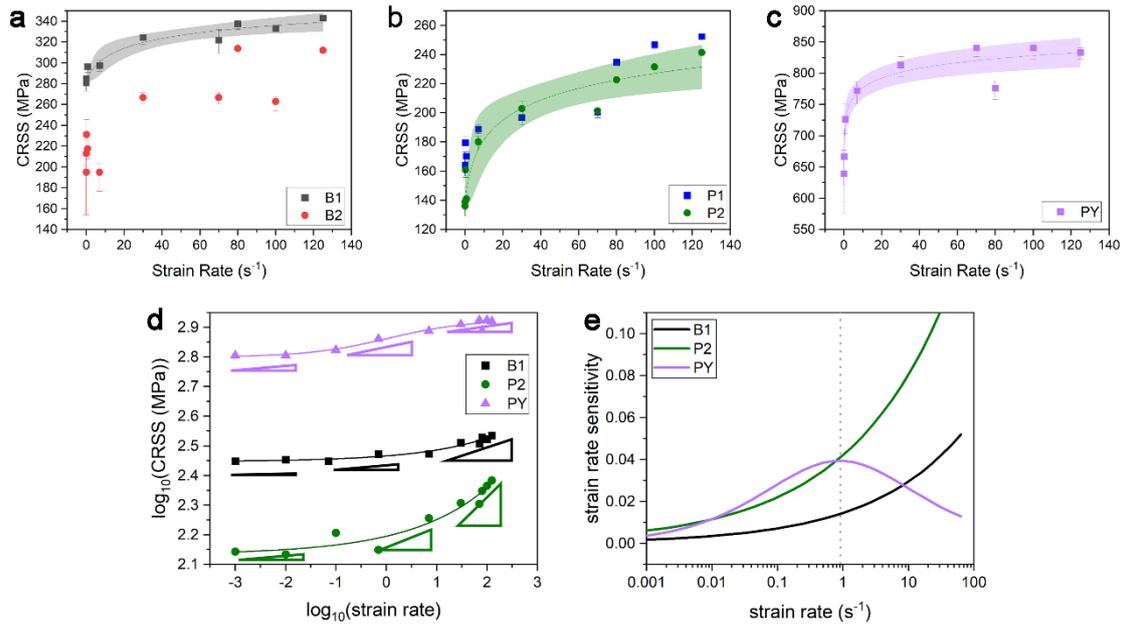

*Figure 11 The CRSS varies with strain rates in the situations with different slip systems (a) B1 & B2, (b) P1 & P2, (c) PY, as fitted to the model as given in Equation 3, with a 95% confidence band indicated with the shaded region. (d) Evaluation of the experimental CRSS vs. strain rate. (in log10-log10 form); (e) The strain rate sensitivity vs. strain rate curve from the slope of fitted curves in (d).*

The CRSS values with respect to different strain rates for different slip systems are plotted using a base 10 logarithmic analysis in Figure 11(d), and the changing slopes of the curves in Figure 11(d) are used to measure changes in the strain rate sensitivity.

At quasistatic rates (i.e. $\log_{10}$ strain rate between -1 to -3) the gradient of the B1 <a> basal slip curve is smaller than both the P2 <a> prismatic and PY <c+a> pyramidal slip. However, there is a significant transition in behavior when transitioning towards higher rates, especially where the strain rate sensitivity of the <a> prismatic slip increases rapidly. This means that strain rate sensitivity vs. strain rate curves of B1 and P2 can be easily fit using an exponential function, but PY seems to have a second plateau at higher rates which requires fitting using a Boltzmann function. Fitting of these functions enables analysis of the gradients analytically, as shown in Figure 11(e), where the slip is plotted as a function of strain rate.

The literature often classifies materials as strain rate sensitive or strain rate insensitive. This can be decided based upon a threshold strain rate sensitivity within a specific loading regime. In this paper, we select a threshold SRS value of 0.04, where if the SRS is smaller than 0.04 the slip system is not considered strain rate sensitive. This threshold value can be used to interpret Figure 11(e) and reveals that for strain rates < 1 s$^{-1}$, the slip systems are not strain-rate sensitive. This regime changes when the strain rate is between 1 s$^{-1}$ and 100 s$^{-1}$ where now only the pyramidal slip system is not strain rate sensitive.

### 4.2. Exploring the model – a comparison of strain rate sensitivity between slip systems

A unified expression among different slip systems to predict the highest SRS value at higher strain rates and the trend of SRS curves, thus the fitting of the CRSS with strain rate was performed using a method developed by Yang *et al.* [40] in Equation 2, where the corresponding single crystal uniaxial loading direction stress σ is assumed to vary as a function of plastic strain rate $\dot{\varepsilon}^p$, Schmid factor $M^S$ and intrinsic material properties:

$$\sigma = \frac{1}{M^S}\left[\frac{kT}{\Delta V^S} sinh^{-1}\left(\frac{\dot{\varepsilon}^p}{\eta \exp\left(-\frac{\Delta F^S}{kT}\right)M^S}\right) + \tau_c^s\right] \qquad (2)$$

where activation energy $\Delta F^s$, activation volume $\Delta V^s$ and slip strength $\tau_c^s$ are the key properties controlling the SRS of each slip system, in which superscript 's' means these parameters are slip-system-dependent.

This can be written into the form of a *sinh$^{-1}$* function of the plastic strain rate $\dot{\varepsilon}^p$ for fitting the experimental data more easily:

$$y = m * sinh^{-1}\left(\frac{x}{\delta * \exp(n)}\right) + A \qquad (3)$$

where $y = \sigma M^S$, $x = \dot{\varepsilon}^p$, $\delta = \eta M^S$, $m = \frac{kT}{\Delta V^s}$, $n = -\frac{\Delta F^s}{kT}$, $A = \tau_c^s$, with all the definitions of symbols listed in Table 4.

| | | <a> basal(B1) | <a> prism(P2) | 1$^{st}$ <c+a> pyramidal (PY) | units |
|---|---|---|---|---|---|
| $k$ | Boltzmann constant | | $1.38 \times 10^{-23}$ | | $J \cdot K^{-1}$ |
| $T$ | temperature | | 298 | | K |
| $\rho_m$ | mobile dislocation density | | 0.01 | | μm$^{-2}$ |
| $\omega$ | dislocation jump frequency | | $1 \times 10^{11}$ | | Hz |
| $b$ | Burgers vector | | $3.23 \times 10^{-10}$ | $6.07 \times 10^{-10}$ | m |
| $\eta$ | $\rho_m \omega (b^S)^2$ | | 104 | 369 | Hz |
| $M^S$ | Schmid factor | | 0.49 | 0.41 | |
| $\delta$ | $\eta M^S$ | 51.2 | 51.6 | 152.0 | Hz |
| $m$ | $kT/\Delta V^s$ | 13.7±3.4 | 20.0±5.2 | 22.2±3.1 | MPa |
| $n$ | $-\Delta F^s/kT$ | -2.4±0.9 | -2.8±1.1 | -8.4±1.4 | |
| $\tau_c^s$ | slip strength = $A$ | 284.6±3.4 | 143.3±6.1 | 636.1±17.5 | MPa |
| $\Delta F^s$ | activation energy = $-nkT$ | 0.1±0.4 | 1.2±0.4 | 3.4±0.6 | $\times 10^{-20}$ J |
| | | 0.06±0.02 | 0.07±0.03 | 0.2±0.1 | eV |
| $\Delta V^s$ | activation volume = $kT/m$ | 3.0±0.7 | 2.1±0.5 | 1.9±0.3 | $\times 10^{-22}$ m$^3$ |
| | | 8.9±2.2 $b_{<a>}^3$ | 6.1±1.6 $b_{<a>}^3$ | 5.5±0.8 $b_{<c+a>}^3$ | |

*Table 4 Fitting parameters for the micropillar compression testing of Zr4 at room temperature.*

Measurements of CRSS as a function of strain rate (shown in Figure 11(a-c)) are used to solve for the unknown values in Equation 3 ($m$, $n$ and $A$) for each slip system independently. Then, the values of $m$, $n$ and $A$ are converted into physically meaningful constants (substituting values between Equation 2 and Equations 3), i.e. the activation volume, $\Delta V^s$, activation energy, $\Delta F^s$, and slip strength, $\tau_c^s$. Uncertainty bounds for each of these constants can be evaluated from the 95% confidence bounds of the model fit, as shown in Figure 11(a-c).

This analysis enables comparison of the (time-independent) slip strength, i.e. the initial critical resolved shear stress $\tau_c^s$ (in Equation 2) for each slip system as reported in Table 4. This enables an analysis of the ratio of the slip strengths between slip systems <a> prismatic: <a> basal: 1st <c+a> pyramidal = 1: 2: 4.6 at quasi static strain rates, and the ratio changes to 1:1.4:3.6 at high strain rates. This ratio is close to the work of Gong *et al.* [36] as reported for commercially pure zirconium.

Furthermore, this analysis enables evaluation of the relative strain rate sensitivity through consideration of the activation volume and activation energy for these slip systems. The activation energy of basal slip ($\Delta F_{basal}^s = 0.06 \pm 0.02$ eV) and prismatic slip ($\Delta F_{primatic}^s = 0.07 \pm 0.03$ eV) are similar, but both of them are smaller than half of the activation energy of pyramidal ($\Delta F_{pyramidal}^s = 0.2 \pm 0.1$ eV). This model is based upon Gibbs's theory [41] of strain rate being controlled by a density of gliding dislocations and developed by Dunne *et al.* [42]. Here, the activation energy is the energy required for the potential dislocation escape to enable glide, which represents the energy barrier that must be overcome for the dislocation to take place. The activation volume describes a pinning of the gliding dislocations within a volume of material, $\Delta V^s$. The model fitting as reported in Table 4 reveals that $\Delta V^S_{basal} > \Delta V^S_{primatic}$, with a ratio of 1.45. Note that this model presents a physically-based prediction of the strain rate sensitivity of deformation within a specific material, where the model is likely useful for materials with similar processing history and strengthening (e.g. the composition of the material, and the presence of other microstructural defects that affect dislocation slip).

When the activation volume ($\Delta V^s$) is smaller, the system is more strain-rate sensitive. Changes in activation energy ($\Delta F^s$) have a more complicated impact. A higher activation energy will make the material more strain rate sensitive at lower strain rates. This motivates a comparison of these competing factors through further exploration of the model.

Many studies in the literature explore strain rate sensitivity (SRS) with a simpler macroscopic approach, by calculating the rate of change of stress and strain rate sensitivity as shown in Equation 4:

$$\text{SRS} = \frac{\partial ln\sigma}{\partial ln\dot{\varepsilon}^P} \qquad (4)$$

This equation can be linked directly to the slip model, via Equation 5 [43]:

$$\text{SRS} = \frac{\dot{\varepsilon}^P}{\sqrt{\left(\eta \exp\left(-\frac{\Delta F^S}{kT}\right)M^S\right)^2 + (\dot{\varepsilon}^P)^2} * \left[\frac{\dot{\varepsilon}^P}{\eta \exp\left(-\frac{\Delta F^S}{kT}\right)M^S} + \frac{\Delta V^S}{kT} * \tau^S_c\right]} \qquad (5)$$

Therefore, the SRS curves for three different slip systems can be plotted as shown in Figure 12.

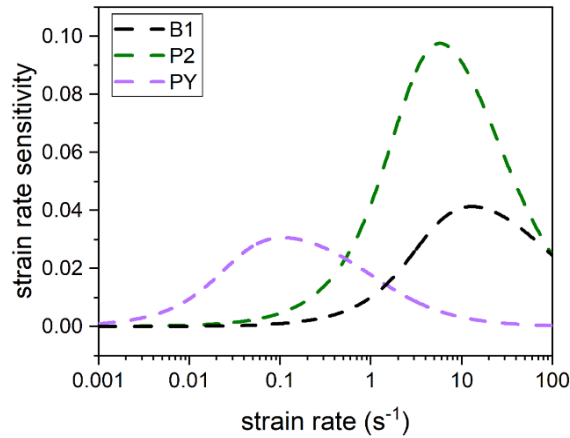

*Figure 12 The strain rate sensitivity vs. strain rate curves from the slip rules model for different slip systems (based on Equation 5 and data in Table 4).*

When Figure 12 is compared with Figure 11(e), similar trends can be found while all curves shift to the right by approximately one order of magnitude, and the largest SRS values are smaller than the highest value in Figure 11(e). The overall trend of the curve rising first and then declining corresponds with the test results [44] and simulations [32] of polycrystalline materials reported in the literature. This is mainly due to the fact that Figure 12 experienced two fitting processes and there is some error in fitting to a specific equation from the slip rule. Despite the presence of such an error, the two plots can corroborate each other, demonstrating the trend of strain rate sensitivity variation for different slip systems on the one hand, and its ability in predicting the trend not yet shown with a strain rate larger than 100 in experiments on the other hand.

The strain rate sensitivity of slip systems is a crucial aspect in understanding the activation and behavior of these systems. In slip processes, approximately 95% of the kinetic energy is partitioned into two primary forms, which are deformation energy and heat. The role of thermal conductivity in heat transfer and its implications on strain rate sensitivity is of particular interest.

At the macroscopic scale, thermal conductivity is generally considered to be a material property. However, understanding the factors that influence strain rate sensitivity requires delving into the microscopic scale where the anisotropy of the thermal properties associated with the orientation of each crystal can be important.

For micropillar compression, as strain rates increase, the time available for heat dissipation from the micropillar to the base material decreases (noting that these tests were performed in a vacuum), resulting in a higher concentration of energy and increasing temperature. This phenomenon can significantly impact the behavior of slip systems under different loading conditions. A local temperature rise during plastic deformation can result in a reduction in CRSS, leading to increased slip localization and the formation of parallel slip bands. The inability to efficiently release the heat generated in these localized regions contributes to the development and intensification of slip bands within the material. This process could be important when scaling these measurements towards larger tests and could motivate even more study.

Furthermore, the transfer of energy during slip is not a one-way process. While the applied loading at the top of the pillar in effect creates an energy wave that moves from the top to the bottom of the micropillars, there is also a feedback where energy comes from the bottom to the top. This bi-directional energy transfer leads to complex interactions between upward and downward energy waves, potentially expanding the effects caused by strain rates. This will be even more important at even higher strain rates (beyond those tested here).

In conclusion, strain rate sensitivity in slip systems is influenced by multiple factors, including thermal conductivity, atomic arrangement, and the presence of defects. The concentration of energy at higher strain rates, along with the complex bi-directional energy transfer within micropillars, makes the study of strain rate sensitivity a compelling area for further research and understanding the fundamental behavior of materials under mechanical deformation.

### 4.3. The appearance of slip systems with different dislocation sources at varying strain rates

Post-test SEM analysis demonstrates that the nature of slip varies with strain rate, as the number and location of the slip traces systematically vary. At higher strain rates, more parallel slip bands appear, and each slip band shows less slip on each slip plane. Some representative post-deformation figures are listed in Figure 13.

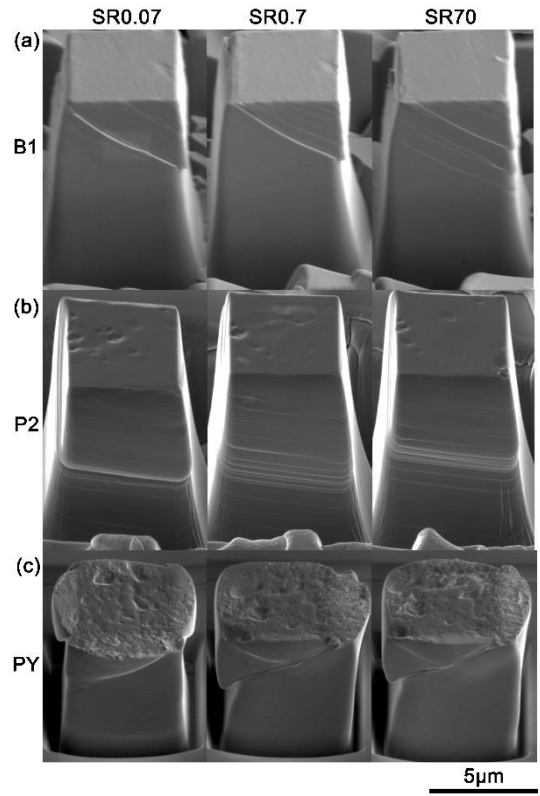

*Figure 13 SE micrographs showing post-deformation micropillars in different grains B1, P2 and PY, compressed with different strain rates. [The SE micrographs of the other 3 sides of the pillar can be found in Supplementary.]*

For <a> basal slip, one major slip trace is typically seen on the side of the pillar (shown in Figure 13(a)) at quasi static strain rates. With a larger strain rate, more parallel slip traces on both sides of the main slip appear, which indicate the increasing number of dislocation source. In contrast, <a> prismatic slip shows a large number of parallel slip traces that cover almost the entire pillar side, especially at higher rates (shown in Figure 13(b)). These observations suggest that at higher strain rates, <a> prism slip operates on multiple parallel slip traces (i.e. easy nucleation in parallel planes) and the dislocation source will evolve from a line to a band.

The 1st <c+a> pyramidal slip system results in large deformations on the upper surface (as shown in Figure 13(c)), and all the deformations are concentrated in the upper part of the micropillar, and this is localised towards the top of the pillar where the shear stress is highest. Once this slip system activates, it causes localised plastic strain at the top of the pillar (leaving a deformed pillar with a 'mushroom' top). This is the main reason which causes the drop in stress when the strain reaches ~0.12 in Figure 10(a). With a higher strain rate, the slip trace will be more obvious, and the edge of the deformed pillar top will be shaper.

### 4.4. The appearance of different slip modes at varying strain rates

B2 and P1 show the activation of more than one slip system with each pillar, and evidence of these slip systems interacting, which can be found in Figure 14.

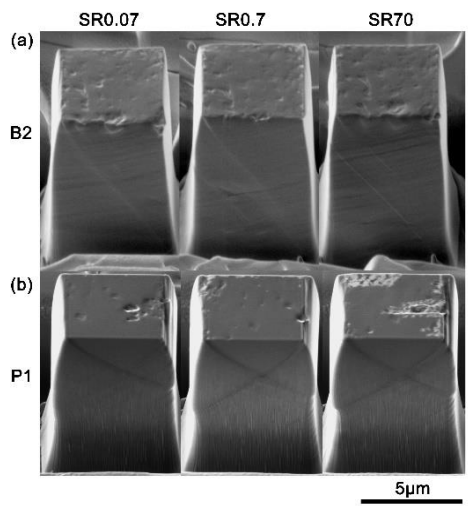

*Figure 14 SE micrographs showing post-deformation micropillars in different grains B2 and P1, compressed with different strain rates. [The SE micrographs of other sides of the pillar can be found in Supplementary.]*

For these cases where multiple slips can operate, activation of the second slip system is controlled by the local stress state at the point when the slip system activates, and this differs from the idealized uniaxial stress state, together with the availability of sources and obstacles for the glide of slip on this second slip system.

It is apparent that with increasing strain rate, the first slip trace is impacted differently from the second slip trace, which is good evidence that strain rate sensitivity is different in the different slip strains. For example, in Figure 14(b), the slip trace from the upper right corner (prismatic slip plane, clarified in Figure 7(c) as grain plane) is affected by strain rate more easily than the slip trace from the upper left corner (1st <a> pyramidal slip plane), although this cannot exclude the effect of obstruction between the different slip systems.

With a simplified analysis of the stress-strain response, deformation of the pillars milled within the B2 and P1 grains can be compared to enable evaluation of the ratio among the $\tau_{CRSS}$ of different slip systems by studying the competition of the 1st, 2nd and 3rd potential activated slip systems and the observed slip traces.

In the post-deformed SEM figures of B2, the basal slip trace with the largest Schmidt factor of 0.45 is clear, and the prismatic slip planes are clearly located on several faces, whose Schmidt factor is 0.3381; while the 2nd largest Schmidt factor for the basal slip system is 0.2749 is not clear. This indicates that the CRSS ratio of basal/prismatic in this situation is between 0.69~1.23 at the quasi-static strain rate, i.e. a potential reduction from the factor of 2 as observed when a single slip is achieved. This motivates further work, outside of the scope of the present study, to use a more sophisticated analysis to explore multiple slip and hardening behaviors.

## 5. Conclusion

In this paper, the quantitative and pictorial differences between the different slip systems in Zr4 at different strain rates are summarized, particularly in the transition area between quasi static and high strain rates (in the range of strain rates $10^{-2}$ s$^{-1}$ and $10^{2}$ s$^{-1}$). Furthermore, the activation of pyramidal slip is clearly achieved and systematically measured as a function of strain rate.

The following conclusions can be made from this work:

- The engineering yield stress increases with a larger strain rate, and the ratio of the CRSS is around <a> prismatic: <a> basal: 1st <c+a> pyramidal = 1: 2: 4.6 at quasi-static strain rates (~1 s$^{-1}$).

- The strain rate sensitivity is different for different slip systems. The prismatic slip system is strain rate sensitive at high strain rates (> 1 s$^{-1}$).

- Activation of the prismatic slip system results in a high density of parallel slip planes which are clearly discrete.

- Activation of <c+a> pyramidal slip results in plastic collapse of the pillar, and results in a 'mushroom' morphology of the deformed pillar.

Future work could focus on these aspects, limited by the constraints of the experimental design due to the initial purpose of the experiment.

- It is difficult to visualize the slip plane or twinning inside the pillar if other materials are interested even after cutting apart due to the redeposition caused by the FIB. Some techniques with higher resolution, such as TEM or HR-EBSD, can be considered to observe the changes in the atomic arrangement around the slip trace.

- *In situ* testing of HSR testing is always a challenge due to the limited scan speed of the SEM. Alternative imaging modes could be explored (e.g. X-ray imaging or perhaps in the TEM).

- Considering the high slip strength and low strain rate sensitivity of grains with a specific orientation, which have the potential to activate the pyramidal slip, this work might be useful in some advanced material manufacturing techniques, such as single-crystal metal foils by contact-free annealing [45], and grain structure control during metal 3D printing [46].

**Data availability**

The data and the supplementary figures in this study are available from Zenodo: [Link to be provided in proof stage]

**Authorship contribution statement**

**Ning Fang:** Conceptualization, Investigation, Methodology, Experiments, Visualization, Writing – original draft & editing. **Yang Liu:** Conceptualization, Writing – review & editing. **Finn Giuliani:** Supervision, Resources, Writing – review & editing. **T. Ben Britton:** Conceptualization, Resources, Writing – review & editing, Supervision.

**Declaration of competing interest**

The authors declare that they have no known competing financial interests or personal relationships that could have appeared to influence the work reported in this paper.

**Acknowledgements**

NF thanks Ruth Birch for useful discussions on sample preparation and helping with the Matlab analysis and Siyang Wang for useful discussions on micropillar compression and the Alemnis stage maintenance. We thank Fionn Dunne for motivating this work and discussions on strain rate sensitivity in hcp metals. We acknowledge funding from EPSRC (EP/S01702X/1).